# Iterative Clustering for Energy-Efficient Large-Scale Tracking Systems

Hesham Alfares[1], Abdulrahman Abu Elkhail[2], Uthman Baroudi[3]*

King Fahd University of Petroleum & Minerals, Dhahran, Saudi Arabia

1. alfares@kfupm.edu.sa, 2. g201536490@kfupm.edu.sa, 3. ubaroudi@kfupm.edu.sa

\* Corresponding author

*Abstract*

A new technique is presented to design energy-efficient large-scale tracking systems based on mobile clustering. The new technique optimizes the formation of mobile clusters to minimize energy consumption in large-scale tracking systems. This technique can be used in large public gatherings with high crowd density and continuous mobility. Utilizing both Bluetooth and Wi-Fi technologies in smart phones, the technique tracks the movement of individuals in a large crowd within a specific area, and monitors their current locations and health conditions. The new system has several advantages, including good positioning accuracy, low energy consumption, short transmission delay, and low signal interference. Two types of interference are reduced: between Bluetooth and Wi-Fi signals, and between different Bluetooth signals. An integer linear programming model is developed to optimize the construction of clusters. In addition, a simulation model is constructed and used to test the new technique under different conditions. The proposed clustering technique shows superior performance according to several evaluation criteria.

## I. INTRODUCTION

In large public events involving large, continuously moving masses of people, it is important to monitor the movement and health conditions of individuals within the crowd. Recent smartphone sets have been used in tracking systems by utilizing their Global Positioning System (GPS) and wireless local area network (WLAN) capabilities for location and communication. In large-scale tracking systems, it is a waste of energy to continuously use the GPS and Wi-Fi features of the smartphones belonging to all individuals in the crowd. This paper presents an energy-saving approach for large-scale tracking systems that limits the use of smartphone's GPS and Wi-Fi features to a few individuals within the crowd. This approach is based on grouping nearby smartphones to form several clusters (groups), where each cluster consists of a cluster head (master) and cluster members (slaves). According to Bluetooth specifications [1], each cluster, also called a *piconet*, can include one master and up to seven slaves. Cluster members communicate locally via low-energy Bluetooth. Only the master nodes uses Wi-Fi to communicate with the back-end server to share current position and health-related data of their cluster members.

This paper presents an efficient heuristic procedure, called the Iterative Clustering Algorithm, to generate near-optimal solutions using a construction process. In addition, a new integer programming model is formulated to optimize cluster formation in large-scale mobile tracking systems. The model determine the number of clusters and designates each cluster's master (head) and slaves (members). The objective of the model is to minimize both the number of clusters and the total distance between cluster masters and members. The ultimate goal is to minimize energy consumption, increase positioning accuracy, and improve transmission quality. Finally, the paper presents a new Matlab Simulink simulation model to evaluate the optimization model's performance under various operating conditions. The objectives of the proposed clustering technique include the following:

(1) **Improving positioning accuracy via short-range radio:** The total distance between masters and slaves is minimized because communicating via short-range radio interfaces such as Bluetooth is more accurate than communicating via long-range radio interfaces.

(2) **Reducing the energy consumption and transmission delay of Bluetooth clusters:** Reducing the total distance between masters and slaves reduces the transmission delay for Bluetooth networks. Furthermore, minimizing the number of clusters minimizes the number of masters that use Wi-Fi, thus minimizing the energy consumption by the masters.

**(3)** **Reducing interference between Bluetooth and Blue-tooth/Wi-Fi:** Minimizing the number of clusters reduces the volume of transmissions, such that interference within the Bluetooth network itself and between Bluetooth and Wi-Fi signals is minimized. In addition, reducing the number of clusters results in reducing channel access congestion.

**(4)** **Maximizing Network lifetime:** Minimizing the number of clusters reduces energy consumption, thus reducing the use of smartphone batteries and maximizing the lifetime of the network.

Subsequent parts of this paper are organized as follows. Section II presents a review of recent relevant literature. Section III provides a general description of the proposed iterative clustering solution process. Section IV presents the integer programming model of the problem. Section V presents optimization and simulation experiments to evaluate the proposed clustering approach. Section VI concludes the paper and provides several directions for future research.

## II. REVIEW OF RELEVANT LITERATURE

The objective of this section is to review and summarize recent relevant techniques for large-scale mobile tracking and positioning systems. The main emphasis is on tracking systems based on clustering techniques, especially those using either Bluetooth or Wi-Fi. The main features of each technique will be summarized, focusing on energy efficiency and low interference of the reviewed techniques. These features are necessary for large-scale tracking applications, especially in high-mobility densely crowded areas.

Weppner et al. [2] developed a system for monitoring crowds in public spaces using mobile devices with a Wi-Fi interface. Using Wi-Fi/Bluetooth interfaces and fixed scanners with directional antennas, the system is used to monitor crowds attending a car manufacturers' exhibition at the Frankfurt Motor Show. The system used a large set of real-life data from 31 scanners, covering an area of 6,000m², 13 business days, and more than 300,000 different mobile devices. The system error has showed to be less than 20% in estimating the crowd density and less than 8m in estimating the positions of individuals. Chen et al. [3] focused on combining smartphone sensors and beacons for accurate indoor localization. The Pedestrian Dead Reckoning (PDR) process is used for localization using smart phone sensors. Since PDR drifts with the walking distance, beacons are introduced to correct the drift using a particle filter.

Experiments show a significant improvement of the localization accuracy with sparse beacons. The main limitation of this approach is that beacons are one-way communication devices.

Kim et al. [4] introduced Bluetooth Low Energy (BLE) mesh approach based on wireless mesh network protocol for BLE. The approach utilizes the broadcasting ability of wireless communications and the results showed decreased energy consumption within the network. However, their study focused only on the navigation aspect and relied on one-way communications by beacons. Mohandes et al. [5] proposed two systems for tracking pilgrims during Hajj. The first system consists of a software that can be downloaded to the mobile phone of every pilgrim. Furthermore, a programmed RFID tag is placed inside the mobile phone. The mobile phone sends the location information through the Internet or SMS to the server for processing and management. The second approach consists of mobile phones carried by pilgrims and a wireless sensor network (WSN) fixed in the region. The WSN communicates the location information of the pilgrims to a server periodically based on pre-set parameters.

Abe et al. [6] developed a tracking system that uses Wi-Fi beacons held by object users and Wi-Fi access points sited widely and densely in a specified area. The positions of object users are estimated based on probe request signals broadcast by the Wi-Fi beacons. The positioning algorithm is based on proximity detection as a function of received signal strength. The experimental results show that the positioning system approach can estimate the positions with approximately 80% accuracy and 2.8s delay time. However, this system is not energy efficient because Wi-Fi signals require higher power.

Conti et al. [7] presented a real-time localization approach using Bluetooth low energy (BLE). An inverse model of the Received Signal Strength Indication (RSSI) and the packet error rate (PER) is used for the estimation of the distance between two BLE devices. The experimental results show that the localization accuracy is significantly improved and the error in the estimation is in general about 1m only. However, the study did not consider tracking purposes but focused only on the localization aspects. Lu et al. [8] proposed indoor positioning technology based on Wi-Fi and the Received-Signal-Strength (RSS) localization method. Their algorithm combines RSS, clustering-based location, and multi-user topology approximation. During the online period, distances between users are measured to reduce the positioning error. During the offline period, the RSS data is collected and the clustering results are corrected. The topology approximation algorithm is used to determine the final localization results. The experimental results show that the localization accuracy is significantly improved. Although a clustering approach is used, it is based on energy-consuming Wi-Fi and focused on estimating the position and network topology.

Alaybeyoglu [9] developed a localization algorithm based on a Sequential Monte Carlo approach in which nodes are able to estimate their speeds and directions for mobile wireless sensor networks. It is assumed that each node's next state is predictable, and hence the particles can be distributed closer to the predicted locations. Therefore, the accuracy of the localization is increased significantly. Lv et al. [10] proposed a localization scheme for mobile wireless sensor networks (WSNs) based on Population Monte Carlo Localization (PMCL) method. A population of probability density functions is used to estimate the distributions of unidentified locations based on a set of observations through an iterative procedure.

Zhang et al. [11] developed two interference-aware approaches. The first approach minimizes interference by skipping the frequencies that are occupied by Wi-Fi. The second approach improves the throughput by restructuring the piconet when the master suffers from interference. However, both approaches focus on a single piconet and do not consider multiple piconets in a bounded area. Yoo and Park [12] presented a distributed clustering approach to reduce power consumption in mobile networks. The approach dynamically adjusts the formation of clusters based on the bandwidth, energy use, and application requirements for each node. Although the approach limits the use of Wi-Fi to cluster heads, it is not designed for tracking purposes. Therefore, it does not consider mobility, location of the nodes, or interference between different signals.

The above literature review shows that that previous methods in general are not fully suitable for large-scale tracking applications with high mobility. This is because they either use high-energy Wi-Fi transmissions, utilize one-way communication devices, ignore the effect of co-existence of hybrid technologies (i.e. Bluetooth and Wi-Fi) on performance, or overlook some practical tracking requirements. In addition, none of the previous papers optimizes the formation of the clusters based on mathematical programming models. This paper presents a new clustering approach that fills these gaps by optimizing Bluetooth clusters while considering mobility, energy consumption, and signal interference of co-existing multiple clusters.

III. DESCRIPTION OF THE CLUSTERING APPROACH

The proposed approach aims to design an energy-efficient system for large-scale wireless tracking applications. To start with, it is required to track and communicate with a large number of nodes in the network (individuals or users, each with their own smart phone). To achieve the goal of lower energy consumption, a cooperative clustering approach is used in which users are divided into small clusters (groups). This is done by grouping neighboring nodes into clusters (piconets), where each cluster has one head (master) and up to 7 members (slaves) [1]. Only the master node of each cluster is responsible for providing location and health data of all cluster members to the back-end server. As shown in Fig. 1, energy-consuming WLAN (Wi-Fi) communication is limited to cluster heads for long-range communication with the server. On the other hand, low-energy Bluetooth is used for short-range communications via personal area networks (PAN) between the masters and the slaves within each cluster. In the proposed approach, Bluetooth version 4.2, also known as Bluetooth Low Energy (BLE), is used for communication between the master and the slaves within each cluster. For optimum performance under this clustering scheme, an integer programing model is used to minimize the number of clusters and hence the number of masters that use Wi-Fi. At the same time, the IP model also minimizes the distances between the masters and the slaves, ensuring faster Bluetooth communications with minimum interference.

(Please place Fig. 1 about here)

After establishing the clusters, communications and sharing of data takes place between the slaves and the master of each cluster via short-range Bluetooth signals. Concurrently, communication takes place via long-range Wi-Fi signals between the masters and the back-end server, where the data is eventually processed and stored. As time goes by, the devices continue to move and to use their battery powers. Therefore, their locations, battery levels, and Wi-Fi connection abilities change. Later on, after a specific short time interval, new clusters are formed with new masters and new sets of members (slaves). Again, the device with the highest battery level and Wi-Fi connection is assigned as master. This periodic change of the master nodes is meant to ensure fair load distribution among the different devices. This process prevents the depletion of individual batteries and maximizes the lifetime of the network.

*Iterative Clustering Algorithm*

The objective of the heuristic iterative clustering algorithm is to design an energy-efficient, low-interference tracking system based on mobile clustering. This system has to be suitable for real-life, large-scale tracking applications with high population density and continuous mobility. To accomplish this objective, the battery levels and Wi-Fi connection availability of each node must be considered, and communications and data exchanges have to be fast and efficient.

Fig. 2 illustrates the steps of the iterative clustering algorithm. The process starts after all the nodes (smart phone devices) in the network are booted up. Immediately, each node will transmit information about its location, battery level, and Wi-Fi connection availability to all neighboring nodes within communication range. All the nodes that have sufficient battery power (above a certain minimum threshold) and Wi-Fi connection availability are eligible to be master nodes. Among those, the node with the highest battery level is selected as the master (head) of the given cluster, and up to 7 nearby nodes within Bluetooth signal range become slaves of this master (cluster members). This process is repeated until each node in the network is designated as either as a master or a slave that belongs to one cluster.

(Please place Fig. 2 about here)

## IV. THE MATHAMATICAL OPTIMIZATION MODEL

The ultimate goal is to design an optimum wireless tracking system based on mobile clustering. In order to meet the practical requirements for applying the system in large-scale environments, energy use must be low, and communication quality must be high. Therefore, the integer programming model presented below aims to optimize the following objectives:

(1)     Minimizing the number of clusters.

(2)     Minimizing the total distance between masters and slaves.

The first objective is pursued because minimizing the number of the clusters is equivalent to minimizing the number of masters that use energy-consuming Wi-Fi. This results in reducing the use of energy and maximizing the lifetime of the network. In addition, minimizing the number of clusters reduces signal transmission traffic, lowering the interference between Bluetooth and Wi-Fi signals and between different Bluetooth signals.

The second objective, which is to minimize the total distance between all masters and their respective slaves, is meant to improve positioning accuracy. For each cluster, the master node is responsible for the positioning information of the cluster members. Minimizing master-slave distances allows for communication via short-range interfaces such as Bluetooth, which is more accurate than using long-range interfaces such as Wi-Fi. Since Bluetooth range is 10m, the maximum error in positioning is ±10m. In addition, shorter distances improve the signal quality and reduce the time delay of Bluetooth transmissions within each cluster.

*A.     Definitions*

Let $i$ = 1 to $n$ denote the slave number, $j$ = 1 to $n$ denote the master number, $C_{ij}$ denote the distance between slave $i$ and master $j$, and $F$ denote the fixed cost per master. Wi-Fi service availability in the user's smartphone ($WF$) is defined as in (1). The user's battery level ($BL$) is defined as in (2). Expressions (3) and (4) define the decision variables, $X_{ij}$ and $Y_j$, which are integer binary variables.

$$WF_j = \begin{cases} 1, & \text{if device } j \text{ has Wi} - \text{Fi connection} \\ 0, & \text{otherwise} \end{cases} \tag{1}$$

$$BL_j = \begin{cases} 1, & \text{if device } j \text{ has battery level} \geq 50\% \\ 0, & \text{otherwise} \end{cases} \tag{2}$$

$$X_{ij} = \begin{cases} 1, & \text{if slave } i \text{ is in the cluster of master } j \\ 0, & \text{otherwise} \end{cases} \tag{3}$$

$$Y_j = \begin{cases} 1, & \text{if node } j \text{ is a master.} \\ 0, & \text{otherwise} \end{cases} \tag{4}$$

The complete integer programming model of the network clustering problem is given by (5). The first expression in (5) is the objective function $Z$, which consists of two terms. The first term is the total distance between masters and slaves, and the second term is the total number of clusters (masters) in the Bluetooth network.

The objective function $Z$ is minimized subject to five sets of constraints. Constraints (I) ensure that every slave has a master. Constraints II limit the cluster size to 8, i.e. 1 master and up to 7 slaves. Constraints III ensure that all cluster members are within the Bluetooth range of their master, i.e. not more than 10m away. Constraints IV ensure that each master node has Wi-Fi connection. Finally, constraints V ensure that a master node's battery level has to be at least 50%. The fixed cost of each master is denoted by $F$ and it is equal to 100.

$$\text{Min } Z = \sum_{i=1}^{n} \sum_{j=1}^{n} \left( C_{ij} X_{ij} \right) + F \sum_{j=1}^{n} Y_j$$

Subject to

I. $$\sum_{j=1}^{n} X_{ij} = 1 \, , i = 1 \dots n$$

II. $$\sum_{i=1}^{n} X_{ij} \leq 8 \, Y_j , j = 1 \dots n$$

(5)

III. $$\sum_{j=1}^{n} C_{ij} X_{ij} \leq 10 \, , i = 1 \dots n$$

IV. $$Y_j \leq W F_j, \quad j = 1 \dots n$$

V. $$Y_j \leq B L_j, \quad j = 1 \dots n$$

## V. NUMERICAL EXPERIMENTS

In this section, the performance of the proposed clustering approach is evaluated by two methods. First, the optimal solutions obtained from the integer programming model are presented. Afterwards, the simulation model results are discussed.

### A. Optimum Solution

To optimally solve the above integer programming model described by (5), the General Algebraic Modeling System (GAMS) was used [13]. Specifically, the mixed integer programming (MIP) feature of GAMS Version 24.3.3 was used. To test the model's performance under varying conditions, the problem was solved assuming four different scenarios.

The first scenario optimizes only the first term in the objective function (minimum total distance). The second scenario optimizes only the second term in the objective function (minimum number of clusters). The third and fourth scenarios simultaneously consider both terms of the objective function. However, the fourth scenario also applies sensitivity analysis by fixing the total number of nodes first to $n = 700$ and then to $n = 800$. This is done while changing the maximum distance between masters and slaves, i.e. changing the right-hand side (RHS) value in constraints III in (5). In addition, sensitivity analysis is applied to both 700 and 800 nodes by changing the fixed cost of each master, $F$, and calculating the optimal value of the number of clusters.

The four above-described scenarios have been studied under the following setup. The dimensions of the area covered by the tracking system are 10m×20m. The optimal objective function values (minimum total distances and number of clusters) have been calculated using GAMS MIP solver with $n$ = 100, 200, 300, 400, 500, 600, 700, and 800 nodes.

Fig. 3 shows the results for scenario 1 (minimizing the total distance). As the number of nodes increases, it can be observed that the total distance between the masters and the slaves is reduced on average. For example, with 100 nodes, the minimum distance is 1m (100.145/100), whereas with 800 nodes it is about 0.6m (477.304/800). Therefore, the clustering approach is effective in reducing the total distances, especially for a large-scale system. A higher accuracy of positioning can be achieved, since short-range radio interfaces are more effective than long-range radio interfaces for localization. Shorter distances also reduce the energy consumption and the transmission delay of Bluetooth networks.

(Please place Fig. 3 about here)

Fig. 4 illustrates the results for scenario 2 (minimizing the total number of clusters). The number of clusters ranges from 13 for 100 nodes to 100 for 800 nodes. For any number of nodes, the cluster size does not exceed 8 nodes, i.e. no more than 7 slaves per master. This small number of clusters is very good for a large-scale system, because there is minimum channel access congestion. Furthermore, interference among Bluetooth signals of different nodes or between Bluetooth and other technologies such as Wi-Fi can be reduced. Lastly, with a minimum number of clusters, Wi-Fi energy consumption by the masters is reduced, thus maximizing the network's lifetime.

(Please place Fig. 4 about here)

Fig. 5 and Fig. 6 display the results for scenario 3, in which the two objectives (total distance and number of clusters) are combined. The value of the fixed cost per master ($F$) is set to 100 to have a reasonable balance between both terms of the objective function. From Fig. 5, it is clear that the number of clusters increases when the number of nodes increases in the model. Moreover,

Fig. 5 shows that for up to 800 nodes, it is still possible to have 8 nodes per cluster. Fig. 6 shows the sum of both terms of the objective function versus the number of nodes. From the figure, it can be concluded that the total minimum distance slightly increases compared to scenario 1. In Fig. 3, the total distance for 100 nodes is equal to 100.145. In Fig. 6, the total distance for 100 nodes is calculated by subtracting the fixed cost of 14 clusters as: $1540.768 - 100 \times 14 = 140.768$.

(Please place Fig. 5 about here)

(Please place Fig. 6 about here)

Figures 7-9 display the results of scenario 4, in which sensitivity analysis is applied to a system of 700 nodes and another of 800 nodes. Fig. 7 shows the optimal number of clusters versus the maximum distance between masters and slaves. This distance, which is the right-hand side value of constraints III in (5), is varied from 2m to 10m. For 700 nodes, the number of the clusters will be minimum when the distance between master $j$ and slave $i$ is equal to 6m, corresponding to 88 clusters. For the case of 800 nodes, the number of clusters remains constant at a value of 100 as the distance between masters and slaves is changed. This shows that the clustering approach is applicable for highly populated areas.

(Please place Fig. 7 about here)

Fig. 8 displays the total distance of the model when the fixed cost per master $F$ is equal to $10^{E}$, where E = 0, 1, 2…, 10. For 700 nodes, the optimal (minimum) total distance is 353m, which is obtained when $F$ is equal to 100 (E = 2). For the case of 800 nodes, the optimal total distance is 559m, which is also obtained when $F$ is equal to 100. These numbers indicate that the clustering approach is well-suited for large-scale tracking applications.

(Please place Fig. 8 about here)

Fig. 9 shows the optimal number of the clusters when the value of fixed cost per master $F$ is equal to $10^{E}$ where E = 0, 1, 2…, 10. For 700 nodes, the optimal (minimum) number of the clusters is 88 clusters, which is obtained when E = 5, or $F = 10^{5}$. For the case of 800 nodes, the optimal number of clusters remains constant at 100 while $F$ is varying.

(Please place Fig. 9 about here)

*B.      Simulation Experimental Setup*

This section presents the results of simulation experiments used to assess the performance of the proposed clustering approach. For this purpose, a simulation model was constructed using MATLAB Simulink, which is a software tool for analyzing the characteristics of Bluetooth and Wi-Fi transmissions. The model analyzes the broadcasting processes of the Bluetooth transmission (transceiver) systems as described in [14-16].

The simulation model is based on the Bluetooth full duplex voice and data transmission model, which is illustrated in Fig. 10 for two Bluetooth devices. The two devices represent a sender node and a receiver node, or alternatively a master and a slave. Transmission between the two devices can be either by data packet type DM1 or by voice packet types HV1, HV2, HV3, and SCORT [17].

The model shown in Fig. 10 allows the performance of the Bluetooth network to be evaluated in the presence of interference. As an interference source, the 802.11 packet block is generated by a separate independent block to be able to measure the interference between Bluetooth and Wi-Fi when they exist in the same area. The Bluetooth uses 79 radio frequency channels in the industrial, scientific, and medical (ISM) radio band, ranging from 2402 MHz to 2480 MHz [1]. In order to be more accurate in the performance assessment, interference must be estimated not only between Bluetooth and Wi-Fi signals, but also between different Bluetooth signals. Therefore, the model was modified by adding the transmitting power signal be able to measure the interference between different Bluetooth signals.

(Please place Fig. 10 about here)

In order to study the performance of Bluetooth clusters in densely populated areas, the model considers $N$ Bluetooth clusters existing together in an area of $10 \times 10$ m$^2$. Therefore, each cluster is subject to interference by $N$-1 other clusters. If several clusters broadcast a message on the same frequency, the sent messages can collide and get lost or distorted. When this happens, the quality of data transmission declines due to interference. Data transmission quality is measured by the frame error rate (FER), which is the proportion of incorrect and missing data out of the total received data. According to Bluetooth standards, all clusters randomly select a channel among 79 possible frequency channels. The model is capable of detecting interference between different Bluetooth signals and also between Bluetooth and Wi-Fi signals [18].

Using the DM1 data packet type, the average Frame Error Rate (FER) per cluster was calculated for each master and slave assuming a different number of clusters. The distance to surrounding clusters was also changed randomly (from 0.1 to 10 meters) 20 times, and the average FER was calculated in order to achieve a 95% confidence interval. It is assumed that the flow data volume of each Bluetooth device is fixed in the cluster with a frame size of 20 bytes (160 bits).

Fig. 11 displays the process of calculating the average frame error rate (FER) for one cluster consisting of one master and one slave that is subject to interference by $N$-1 other clusters. The single-slave cluster is sufficient to represent a fully loaded seven-slave cluster, as time division multiple access (TDMA) is used to manage the channel access and one user is active in each time slot. From Fig. 11, it can be observed that the average FER of the master is greater than that of the slave. As expected, the average FER increases as the number of the clusters increases, leading to a higher degree of interference. This is the main reason for making the minimum number of clusters a main objective in the proposed clustering approach. By minimizing the number of clusters, the channel access congestion is reduced, and consequently the interference is significantly lowered between Bluetooth and Wi-Fi signals and also between different Bluetooth signals.

(Please place Fig. 11 about here)

*C.    Performance Metrics*

The performance of three methods was compared for solving large-scale wireless tracking systems. The first method is the direct approach, in which the nodes are not clustered, but each node communicates with the server directly using its Wi-Fi and GPS connection. The second method is the iterative clustering algorithm described in Section III. The third method is the optimal GAMS solution of the integer programming model presented in section IV and specified by (5). Matlab was used to evaluate the performance of these three methods.

The same experimental setup was used for the three methods. Each node can send data traffic at a rate of 1,000 kbps at frame sizes up to 20 bytes, which is sufficient for health information messages. The input parameter values specified by Yoo and Park [12] were used to determine the Bluetooth and Wi-Fi energy consumption. In order to achieve 95% confidence interval, each simulation experiment was repeated 10 times using different random values.

The total energy consumption and throughput in each run were calculated for a different number of nodes, using the following equations.

$$TE = TE_{CH} + TE_{CM} + TE_{idle} \qquad (6)$$

$$TP = R \times CS \times FS \times Pc \qquad (7)$$

$$Efficiency = Throughput \ / \ TE \qquad (8)$$

In the above equations, $TE$ is the total energy consumption of all nodes, which is the sum of energy consumption by cluster heads $TE_{CH}$, energy consumption by cluster members $TE_{CM}$, and energy consumption by idle nodes $TE_{idle}$.

$Throughput$ is defined as the total number of successfully received bits, $R$ is the number of rounds (i.e. time intervals), $CS$ is the cluster size, $FS$ is the frame size, and $Pc$ is the frame correction rate, where ($Pc = 1 − FER$). Finally, $Efficiency$ is defined as the $Throughput$ divided by the total energy consumption.

Fig. 12 shows the average throughput of the iterative clustering approach for different values of the total number of nodes. As expected, the average $Throughput$ increases as the number of nodes increases, since more data is sent and received through the network.

(Please place Fig. 12 about here)

*D.    Simulation Results*

Under scenario 3, energy consumption was compared for the direct approach, the iterative clustering approach, and the optimal GAMS solution. Considering different values for the total number of nodes, the average values of energy consumption for each method are shown in Fig. 13 and Table 1. From Fig. 13, it is observed that the energy usage of the iterative clustering approach is very close to the minimum energy usage of the optimal IP solution obtained by GAMS. Moreover, the energy needs of the iterative clustering approach become closer to optimality as the number of nodes increases. This fact is obvious from Table 1, which shows a difference of 1.8% in energy consumption between the performance of the iterative clustering approach and the optimal solution with 100 nodes, and a difference of zero with 800 nodes. This comparison indicates that the proposed iterative clustering algorithm provides near-optimum solutions for large-scale tracking problems.

For the direct approach, the total energy increases as the number of nodes increases. This is because the direct approach requires each node to use Wi-Fi and GPS for transmission of the data to the back-end server. Since all nodes transmit data over long-range, the direct approach consumes more energy than the proposed clustering approach. As observed from Table 1, the energy consumption of the direct approach is 377.8% higher than the optimal consumption specified by GAMS when the number of nodes is equal to 100, and 409.4% higher when the number of nodes is equal to 800. Clearly, the direct approach is not a practical solution method for large-scale high-mobility tracking systems.

(Please place Fig. 13 about here)
(Please place Table 1 about here)

The average energy efficiency values are shown in Fig. 14 and Table 2 for the direct approach and the iterative clustering approach assuming different values for the total number of nodes. For the clustering approach, these values show that efficiency packet per Joule slightly varies with the change in the number of nodes. For the direct approach, however, the average efficiency packet per Joule remains constant as the number of nodes varies. This is expected because each node in the direct approach uses Wi-Fi and GPS to transmit data directly to the server. Therefore, the average energy efficiency per node remains the same regardless of the number of nodes.

(Please place Fig. 14 about here)
(Please place Table 2 about here)

## VI. CONCLUSIONS

A new technique has been presented for the optimum design of energy-efficient large-scale mobile wireless tracking systems. This technique minimizes energy consumption in the system by forming mobile clusters to avoid high-energy, long-range direct communication between each node and the server. Within each cluster, the nodes communicate using low-energy, short-range Bluetooth signals. Only one master node in each cluster uses long-range Wi-Fi transmissions to provide location and health data of all cluster members to the server. In order to optimize performance, the proposed clustering algorithm also minimizes the number of clusters and the total distance between master nodes and member nodes. By minimizing the distances and the number of clusters, the proposed technique achieves several desirable objectives. These objectives include lower energy consumption, transmission delay, and signal interference. In addition, the proposed technique provides for higher positioning accuracy and longer network lifetime. Results of simulation experiments show that the iterative clustering algorithm succeeds in producing near-optimal solutions that achieve these objectives. This means that the new clustering technique is suitable for real-life applications in large-scale mobile tracking systems.

Based on the optimization model and the iterative clustering heuristic algorithm presented in this paper, there several directions for future research aimed at designing energy-efficient large-scale tracking systems. For example, in addition to Bluetooth, other technologies and devices such as sensors, beacons, RFID tags, and antennas could be used to improve the performance of wireless tracking systems. Another interesting extension is to consider the movement of individuals (nodes) to be not completely random, but to be in the general direction of a set of destinations, or to be affected by the paths, obstacles and general layout of the area. A third extension is to consider other options for reducing interference, such as imposing a minimum separation distance between different master nodes.

**ACKNOWLEGMENTS**

The authors Abdulrahman Abu Elkhail and Uthman Baroudi would like to acknowledge the support provided by the Deanship of Scientific Research (DSR) at King Fahd University of Petroleum and Minerals, under the grant RG1424-1.

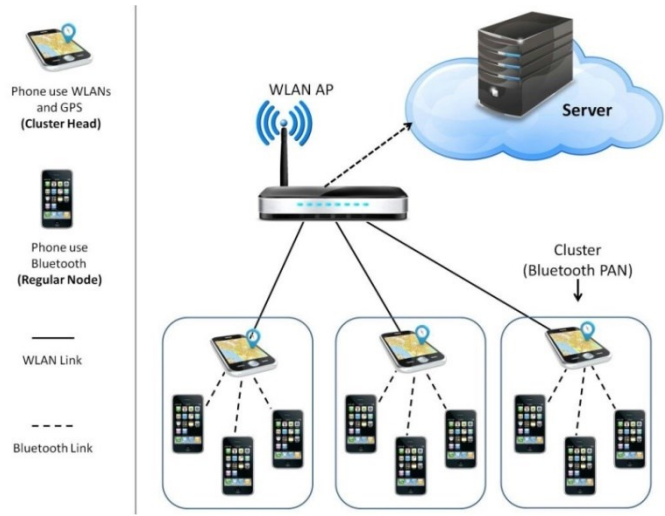

Fig. 1. Proposed clustering approach: slaves use Bluetooth while masters use Wi-Fi.

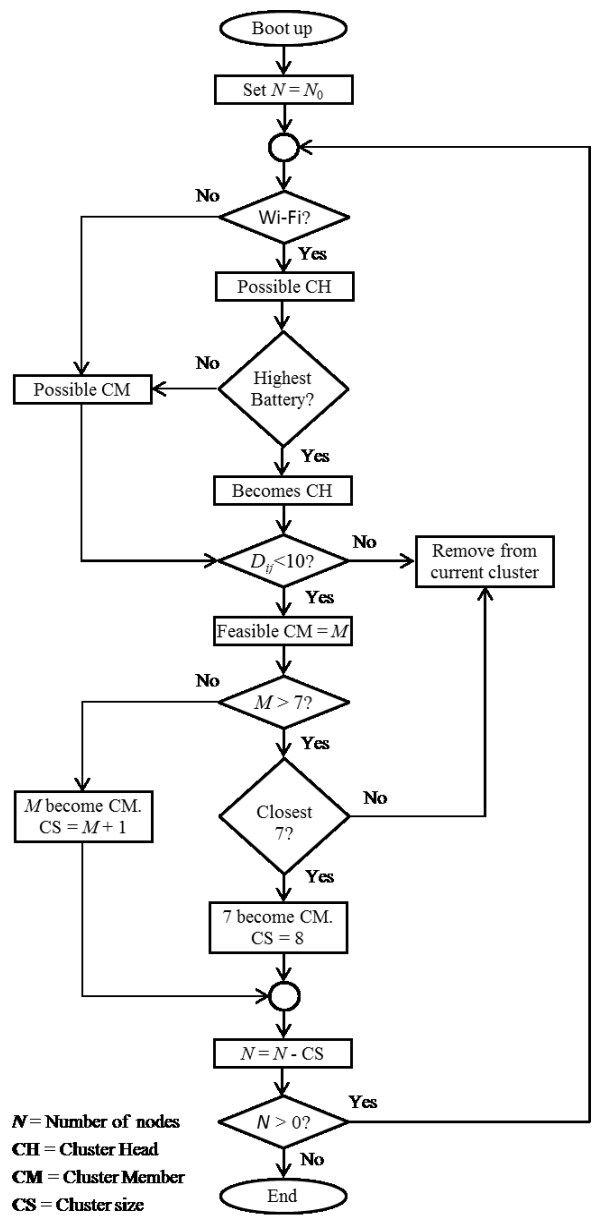

Fig. 2.   Flow Chart of the Iterative Clustering Algorithm

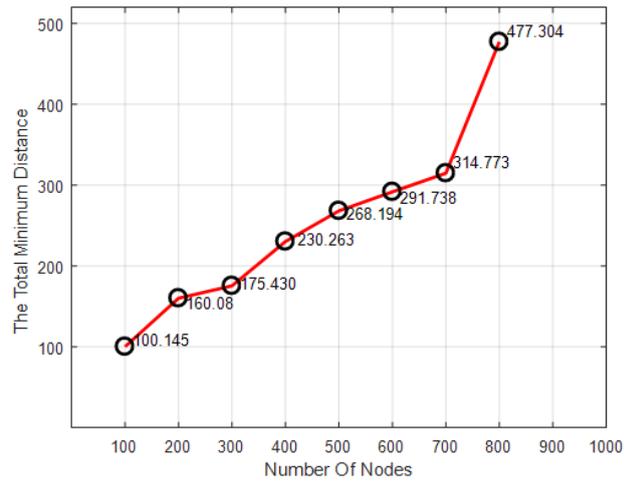

Fig. 3. The total minimum distance (for scenario 1)

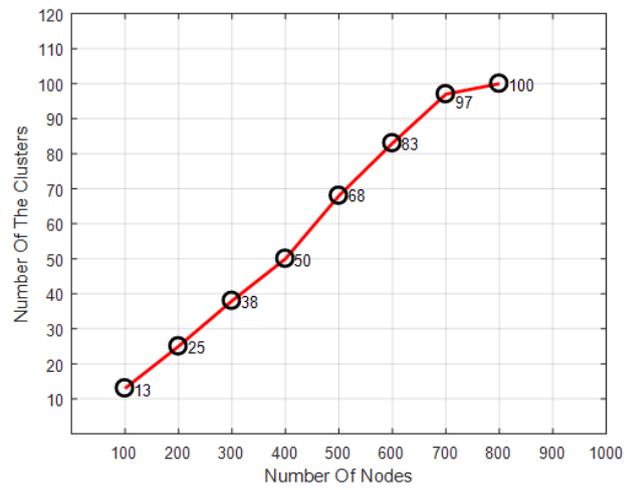

Fig. 4. Optimal number of the clusters (for scenario 2).

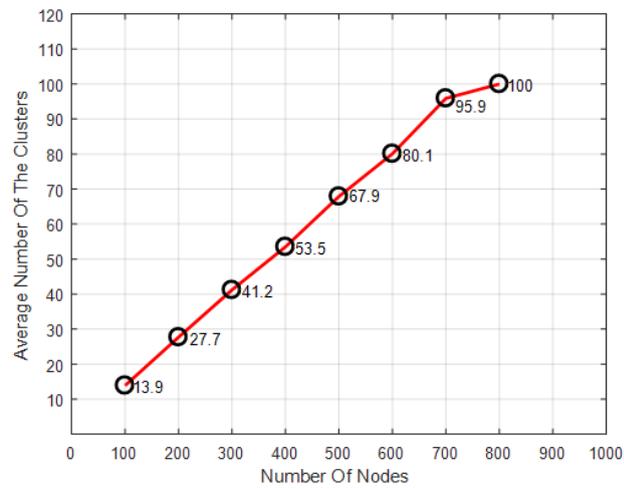

Fig. 5. Optimal number of clusters (for scenario 3).

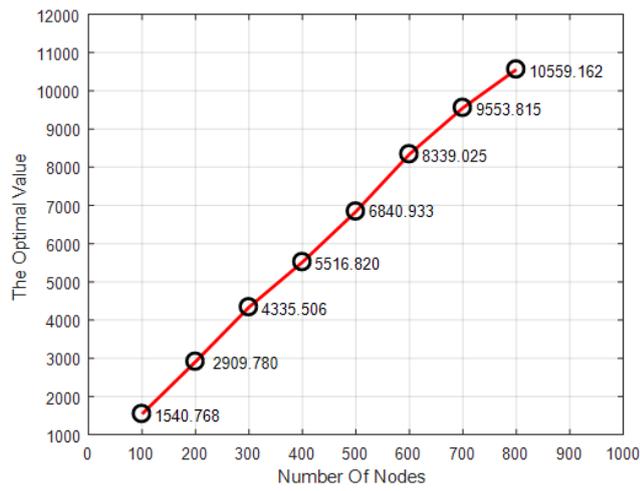

Fig. 6. The optimal objective function value of model (for scenario 3)

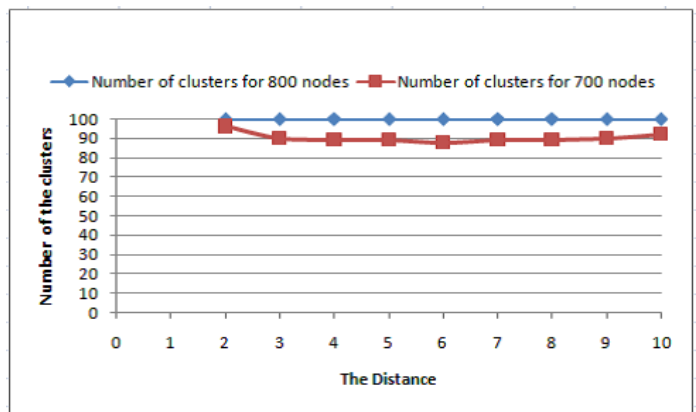

Fig. 7. Optimal number of clusters versus the maximum allowable distance between a master and slave nodes.

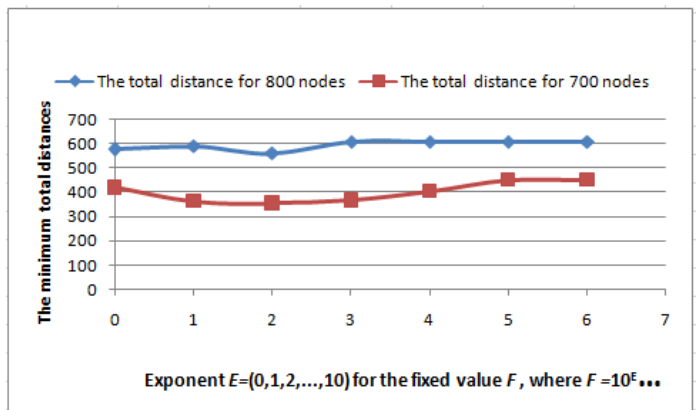

Fig. 8. The total distance when changing F.

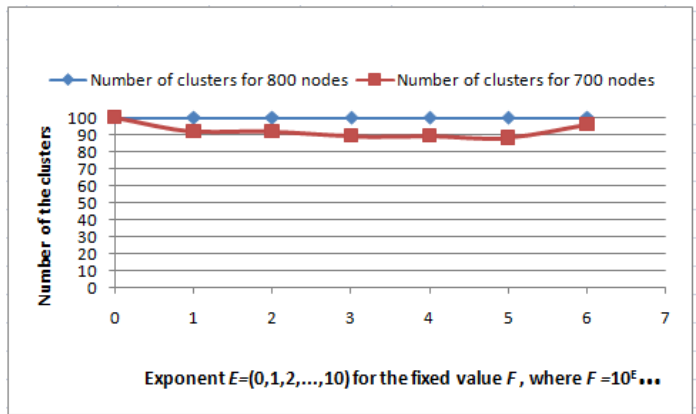

Fig. 9. Optimal number of clusters when changing F.

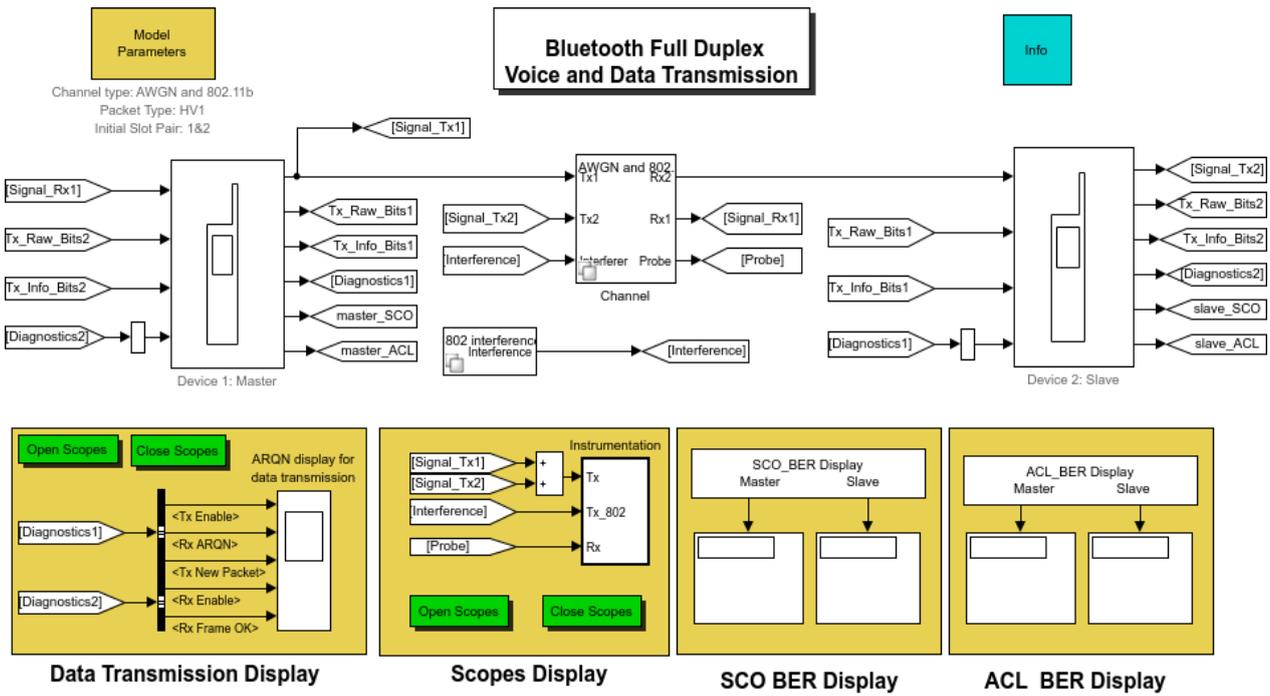

Fig. 10. Bluetooth Full Duplex Voice and Data Transmission model.

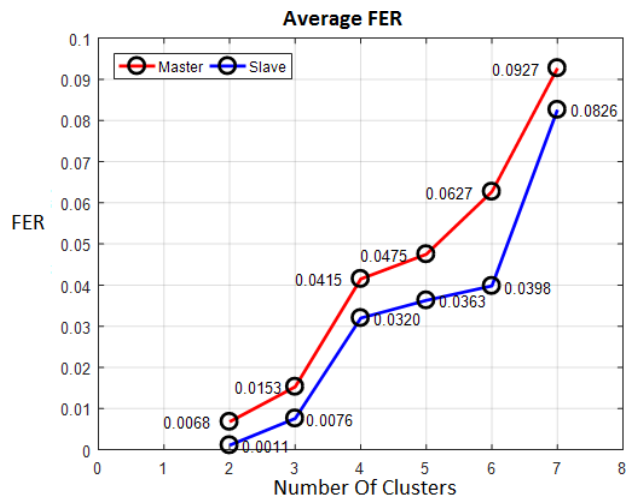

Fig. 11. Average frame error rate for multiple Bluetooth coexisting piconets.

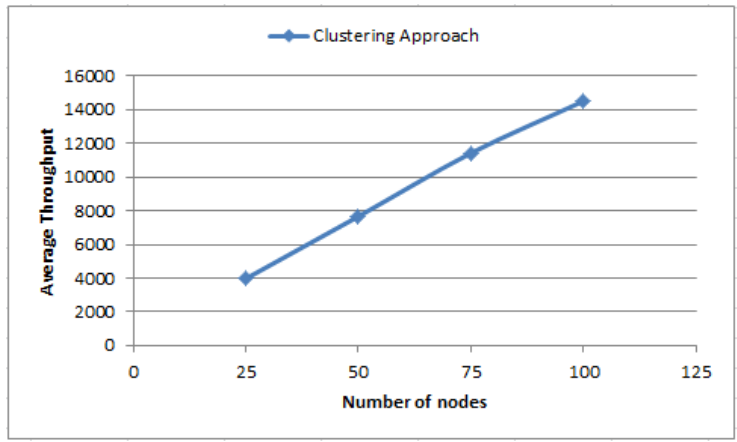

Fig. 12. Average throughput of the iterative clustering approach.

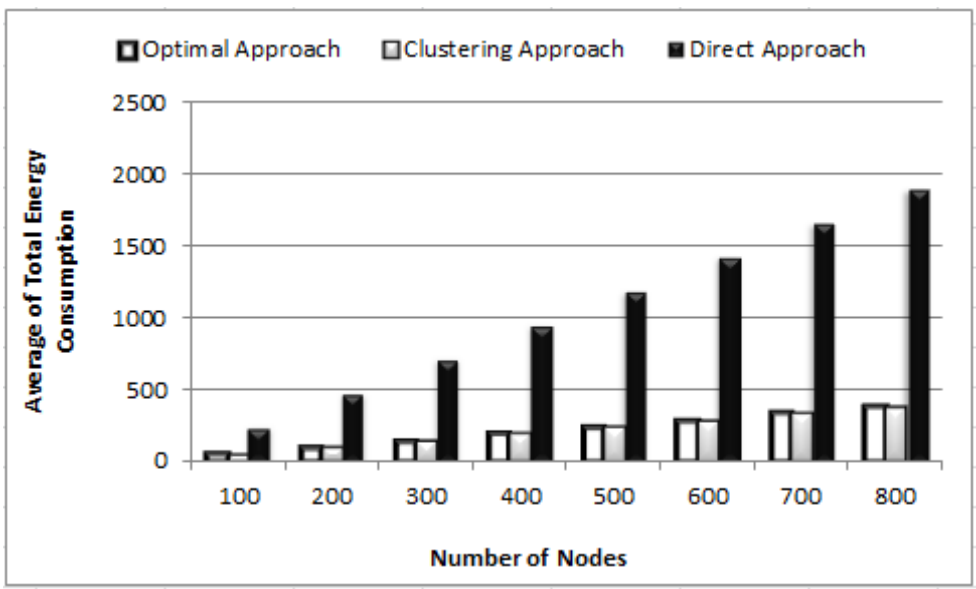

Fig. 13. Comparison of the average Energy Consumption under scenario 3.

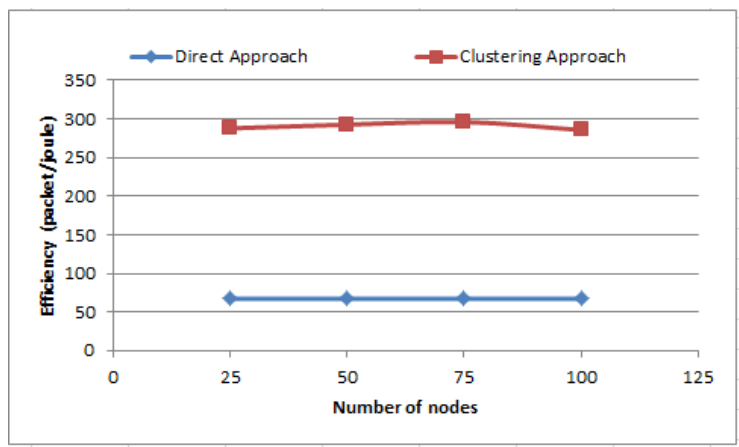

Fig. 14. Comparing the energy efficiency of direct approach and clustering approach.

Table 1. Comparison of total energy consumption in Joules for three solution methods.

| #of nodes | Iterative Clustering Approach | Direct Approach | Optimal Approach GAMS | Comparison vs GMAS | |
|---|---|---|---|---|---|
| | | | | Iterative Clustering Approach | Direct Approach |
| 100 | 50.8 | 238.41 | 49.9 | 1.80% | 377.80% |
| 200 | 97.9 | 476.82 | 96.4 | 1.60% | 394.60% |
| 300 | 145.9 | 715.23 | 144.2 | 1.20% | 395.90% |
| 400 | 191.2 | 953.64 | 189.5 | 0.90% | 403.20% |
| 500 | 237.8 | 1192.05 | 236.6 | 0.50% | 403.80% |
| 600 | 284.1 | 1430.46 | 283.2 | 0.30% | 405.10% |
| 700 | 330.1 | 1668.87 | 329.7 | 0.10% | 406.20% |
| 800 | 374.4 | 1907.28 | 374.4 | 0% | 409.40% |

Table 2. Comparison of energy and efficiency of the direct approach and the clustering approach

| #of nodes | Throughput of (bps) | | Energy Consumption (Joule) | | Efficiency of (packet/joule) | |
|---|---|---|---|---|---|---|
| | Clustering Approach | Direct Approach | Clustering Approach | Direct Approach | Clustering Approach | Direct Approach |
| 25 | 3972.8 | 4000 | 13.8 | 59.6 | 287.9 | 67.1 |
| 50 | 7668 | 8000 | 26.2 | 119.2 | 292.7 | 67.1 |
| 75 | 11430 | 12000 | 38.6 | 178.8 | 296.1 | 67.1 |
| 100 | 14516.8 | 16000 | 50.8 | 238.41 | 285.8 | 67.1 |

**REFERENCES**

[1] Bluetooth SIG (2017). Bluetooth Specifications, Bluetooth Technology Website (https://www.bluetooth.com/).

[2] Weppner, J.; Bischke, B.; & Lukowicz, P.: Monitoring crowd condition in public spaces by tracking mobile consumer devices with wifi interface. In Proceedings of the 2016 ACM International Joint Conference on Pervasive and Ubiquitous Computing, 1363-1371 (2016)

[3] Chen, Z.; Zhu, Q.; Jiang, H., & Soh, Y. C.: Indoor localization using smartphone sensors and iBeacons. In IEEE 10th Conference on Industrial Electronics and Applications (ICIEA), 1723-1728 (2015)

[4] Kim, H. S.; Lee, J.; & Jang, J. W.: Blemesh: A wireless mesh network protocol for bluetooth low energy devices. In 3rd International Conference on Future Internet of Things and Cloud (FiCloud), 558-563 (2015)

[5] Mohandes, M.; Haleem, M. A.; Kousa, M.; & Balakrishnan, K.: Pilgrim tracking and identification using wireless sensor networks and GPS in a mobile phone. Arabian Journal for Science and Engineering, 38, 2135-2141 (2013)

[6] Abe, R.; Shimamura, J.; Hayata, K.; Togashi, H.; & Furukawa, H.: Network-Based Pedestrian Tracking System with Densely Placed Wireless Access Points. In: Information Search, Integration, and Personlization, pp. 82-96, Springer (2017)

[7] Conti, M.: Real Time Localization Using Bluetooth Low Energy. In International Conference on Bioinformatics and Biomedical Engineering, 584-595 (2017)

[8] Lu, X.; Wang, J.; Zhang, Z.; Bian, H.; & Yang, E.: WIFI-Based Indoor Positioning System with Twice Clustering and Multi-user Topology Approximation Algorithm. In International Conference on Geo-Informatics in Resource Management and Sustainable Ecosystems, 265-272 (2016)

[9] Alaybeyoglu, A.: An efficient Monte Carlo-based localization algorithm for mobile wireless sensor networks. Arabian Journal for Science and Engineering, 40, 1375-1384 (2015)

[10] Lv, C.; Zhu, J.; & Tao, Z.: An Improved Localization Scheme Based on PMCL Method for Large-Scale Mobile Wireless Aquaculture Sensor Networks. Arabian Journal for Science and Engineering, 43, 1033-1052 (2018)

[11] Zhang, Q.; Chen, G.; Zhao, L.; & Chang, C. Y.: Piconet construction and restructuring mechanisms for interference avoiding in bluetooth PANs. Journal of Network and Computer

Applications, 75, 89-100 (2016)

[12] Yoo, J. W.; & Park, K. H.: A cooperative clustering protocol for energy saving of mobile devices with wlan and bluetooth interfaces. IEEE Transactions on Mobile Computing, 10, 491-504 (2011)

[13] GAMS Software GmbH (2017). GAMS Specifications, GAMS Website, (https://www.gams.com/).

[14] Golmie, N.; Van Dyck, R. E.; Soltanian, A.; Tonnerre, A.; & Rebala, O. Interference evaluation of Bluetooth and IEEE 802.11 b systems. Wireless Networks, 9, 201-211 (2003)

[15] Santivanez, C.; Ramanathan, R.; Partridge, C.; Krishnan, R.; Condell, M.; & Polit, S.: Opportunistic spectrum access: Challenges, architecture, protocols. In Proceedings of the 2nd Annual International Workshop on Wireless Internet, 13- (2006)

[16] Hu, Z.; Susitaival, R.; Chen, Z.; Fu, I. K.; Dayal, P.; & Baghel, S. K.: Interference avoidance for in-device coexistence in 3GPP LTE-advanced: Challenges and solutions. IEEE Communications Magazine, 5, 60-67 (2012)

[17] Mathew, A.; Chandrababu, N.; Elleithy, K.; & Rizvi, S.: IEEE 802.11 & Bluetooth interference: simulation and coexistence. In Seventh Annual Communication Networks and Services Research Conference (CNSR'09). 217-223 (2009)

[18] Chek, M. C. H.; & Kwok, Y. K.: Design and evaluation of practical coexistence management schemes for Bluetooth and IEEE 802.11 b systems. Computer Networks, 51, 2086-2103 (2007)